\documentclass[%
 reprint,
 amsmath,amssymb,
 aps,
]{revtex4-1}
\usepackage{graphicx}
\usepackage{tabularx}
\usepackage{dcolumn}
\usepackage{bm}
\usepackage{braket}
\usepackage{verbatim}
\usepackage{textgreek}
\usepackage{hyperref}
\usepackage{float}
\usepackage{xcolor}
\usepackage[utf8]{inputenc}
\usepackage[export]{adjustbox}

\begin{document}

\title{Stroboscopic detection of multi-colloidal hydrodynamics using time-multiplexed optical tweezers}
\author{Thomas F. Dixon}
 \email{thomas.dixon@unsw.edu.au}
\author{Peter J. Reece}%
\affiliation{%
 School of Physics, University of New South Wales, Sydney, Australia
}%

\date{\today}

\begin{abstract}
Time-division multiplexing presents an attractive opportunity to probe multi-colloidal interactions in optical traps at short time-scales. In this paper, we demonstrate a stroboscopic system capable of arbitrary control of multiple trapped colloids with sensing at kHz rates and validate it using several simple multi-colloidal experiments. We expect this methodology will be of benefit in the study of group colloidal hydrodynamics and systems of active colloids, particularly where a temporal sensitivity beyond that of camera-based position sensing is required. In addition, our multiplexing enables \textit{in situ} calibration that is robust to environmental anomalies, shape distortions of colloids and scattering interference from other particles.
\end{abstract}

\pacs{Valid PACS appear here}
\maketitle


\section{Introduction}
Investigations of colloidal hydrodynamics using optical tweezers yield significant advances to understanding of biological processes \cite{jones2015optical} and the behaviour of colloidal ensembles \cite{lintuvuori2010colloids}. Such investigations demand precise manipulation and monitoring of multiple trapped colloids, which is often difficult to achieve for large arrays of particles \cite{Ott2014simultaneous}. In addition, direct hydrodynamic confinement and manipulation of non-trappable colloids via optical tweezers has been recently demonstrated using trapped colloidal spheres \cite{butaite2018hydrodynamic} and specialised rotors \cite{Butaite2019indirect} which offer promising new avenues of micro-manipulation and assembly but present a methodological challenge to fully characterise.

Passive configurations of optical tweezers, in which the trap positions are fixed, find extensive use in investigation of colloidal crystals --- assemblies of colloids that hydrodynamically organize into long-range-ordered crystals --- \cite{pieranski1983colloidal} by studying the colloid-colloid interactions that give rise to crystalisation \cite{Meiners}\cite{grier1997optical}, or through the manipulation of elements within a colloidal crystal itself \cite{pertsinidis2001diffusion}\cite{pertsinidis2001equilibrium}. Further sensing applications include study of sedimentation \cite{palacci2010sedimentation} and microrheology \cite{Meyer2006laser}\cite{lintuvuori2010colloids} --- the measurement / mapping of viscosity, density and other fluid properties at the micron scale \cite{mackintosh1999microrheology}. 

The time-scales of some of these hydrodynamic interactions are quite short \cite{Meiners}\cite{Reichert2006}, and as such poly-colloidal experiments typically rely on photodiode detection \cite{gittes1998interference}\cite{kreiserman2019decoupling}, which has high temporal and spatial sensitivity \cite{Meiners}. However, photodiode detection for three or more particles is highly impractical \cite{Ott2014simultaneous}, as orthogonal polarisations can no longer be used to minimise interference between the traps \cite{visscher1996construction}. Most investigations of three or more statically trapped particles are performed using Holographic Optical Tweezers (HOTs) \cite{dufresne1998optical}\cite{grier2003revolution}, restricting observations to longer timescale interactions suitable for video tracking \cite{polin2006anomalous}\cite{di2007eigenmodes}. A secondary method is to utilise the fast switching of a beam-steering device like an Accousto Optic Deflector (AOD) to timeshare a single trap into multiple positions\cite{visscher1996construction}. The most pertinent innovation on this front was presented by Ruh \textit{et al.} \cite{ruh2011fast} in 2011, who were able to maintain photodiode compatibility during trapping of multiple particles using a fast switching AOD to timeshare one optical trap into an array of 9 positions (based on an earlier project by the same group \cite{speidel2009interferometric}). As the laser is only active on one particle at a time, the photodiode signal is relatively unperturbed and can be used for position tracking, achieving tracking sensitivity of 1-5 nm with a time resolution of 11 kHz. 

In addition to their use as analogues for passive hydrodynamic processes, optical tweezers are able to simulate and characterise active colloidal systems that closely mimic biological systems \cite{kotar2010hydrodynamic}. By investigating the complex hydrodynamics of such systems, researchers are able to gain insights into hydrodynamically sensitive biological processes such as protein transport \cite{brune1994hydrodynamic}, bio-molecular diffusion \cite{ando2010crowding} and the synchronistic behaviour of motile cilia \cite{bruot2012driving}\cite{kotar2013optimal} (organelles responsible for --- among other things --- removal of harmful material from the lungs \cite{shah2009motile}). Active colloidal optical tweezers also have applications in microrheology \cite{HoughLA2002Cmot}\cite{lintuvuori2010colloids}, where the fluid properties are able to be probed more sensitively by observing particle drag or other dynamic effects \cite{HoughLA2002Cmot}.  The driven oscillation of these trapped colloids is typically accomplished using fast beam-steering via galvano-mirrors \cite{svoboda1994biological}\cite{HoughLA2002Cmot} or AODs \cite{visscher1996construction}\cite{mellor2005probing}, as HOTs possess too low a refresh rate to drive frequencies above a few Hz \cite{grier2003revolution}. These pseudo-active colloidal systems are highly relevant to biological sciences \cite{damet2012hydrodynamically}, but have not yet been implemented alongside photodiode detection, instead requiring the use of video tracking \cite{bruot2012driving}\cite{cicuta2012patterns} which limits the investigations possible to those considering lower-frequency interactions within the observation capacity of conventional CCD/CMOS systems. To probe driven behaviour at short timescales / periods of oscillation, experimenters can use only simplified two-trap setups compatible with photodiode detection \cite{HoughLA2002Cmot}. 

In this paper, we demonstrate a time-sharing methodology that is suitable for trapping, driving and detection of multiple colloids and demonstrate applications for both passive and active colloidal systems. In a principle similar to that of Ruh \textit{et al.} \cite{ruh2011fast}, we strobe the trap position at a fast rate relative to the diffusion time of the colloids, allowing the particles to be localised without the need for continuous power \cite{guilford2004creating} and their behaviour is analogous to a continuous power system with a reduced laser intensity \cite{capitanio2007continuous}. Active colloidal driving is achieved through successive displacement of optical traps. In addition, our system performs a computationally simple \textit{in situ} detector calibration for each trapped object, and is therefore robust to asymmetries between particles and localised environmental factors, while yielding a very straightforward  output that is directly interpretable in terms of particle trajectories. Importantly, we maintain compatibility with photodiode-based detection and are therefore able to probe interactions and drive oscillations at kHz frequencies.
\section{Methodology}
\begin{figure}
\centering
\includegraphics[scale=0.26]{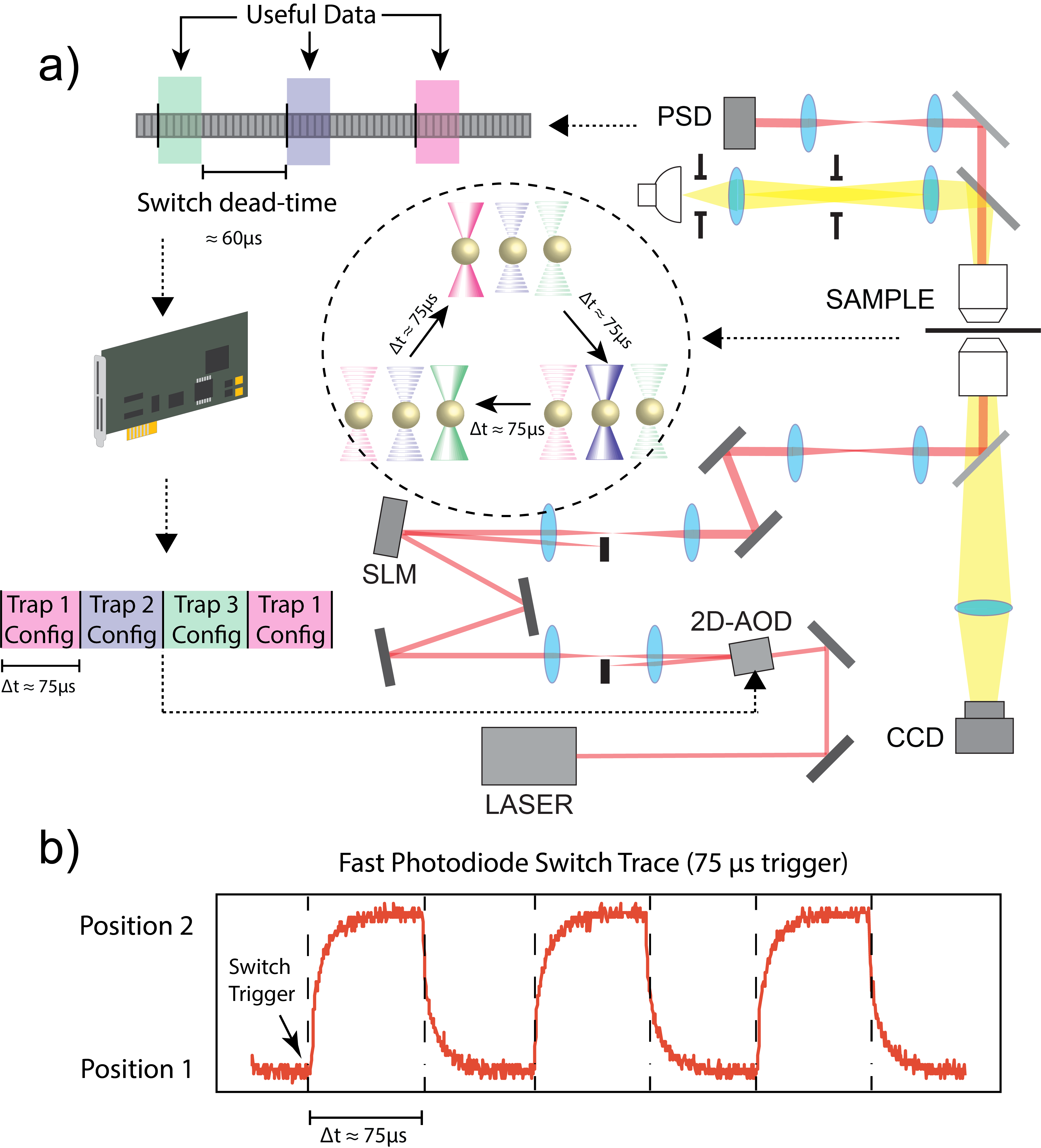}
\caption{\textbf{a)} Apparatus setup: A 1064nm Laser is passed through a two axis Accousto Optic Deflector (2D-AOD), which deflects the laser by a precise angle. The laser is imaged onto a Spatial Light Modulator (SLM) for aberration correction, before being focused via a microscope objective into a sample of 1um polystyrene spheres suspended in water. A Position Sensitive Diode (PSD) positioned at the back focal plane of the condenser detects trapped particle position via back focal plane interferometry. Imaging of the trapped beads (for user operation) is performed via a CCD and bright field illumination. The FPGA drives fast trap strobing from a series of up to 1000 stored configurations by controlling the laser deflection form the AOD via a Digital Frequency Synthesiser (DFS). The PSD is addressed at 200 kHz to yield positional data, from which usable data are extracted and de-multiplexed. There is approximately 75 ms switching time between each trap position (see insert).\textbf{ b)} A switch rate of 75 ms was chosen due to an approximately 60 ms rise/fall time between trap positions. This 'dead-time' is resultant from a combination of sensor delay and slowness in the DFS/AOD combination. }
\label{fig:apparatus}
\end{figure}

The optical trapping setup for these experiments is outlined in Figure \ref{fig:apparatus} a). Trapping of two or more simultaneous beads is achieved through stroboscopic switching of a single optical trap. From a user-defined array of desired trap positions, a Field-Programmable Gate Array (FPGA, National Instruments PCIe-7852R) generates a sequential series of commands to be sent to two digitally addressed Digital Frequency Synthesisers (DFS, Gooch \& Housego, 64020-200-2ADMDFS-A). These DFSs interface with a two-axis Accousto-Optic Deflector (2D-AOD, Gooch \& Housego 45035 AOBD), controlling the power and incident angle of a linearly polarised 1064nm laser (Laser Quantum, IR Ventus), which has been abberation-corrected by a Spatial Light Modulator (SLM, Hammamatsu LCOS-SLM
x10468-03). By rapidly re-addressing the DFS, the FPGA is able to quickly switch between an array of trap positions and powers. The laser is incident on the back aperture of a 1.3 NA microscope objective (Nikon CFI Plan Fluor 100x) which focuses the laser light into a sample chamber containing 1 \textmu m polymer colloids (Thermo Fischer Scientific) suspended in water. A Position Sensitive Diode (PSD, Pacific Sensor DL16-7PCBA) captures the back focal plane interference pattern imaged from a 0.65 NA condenser objective (Nikon Plan Fluor ELWD 40x), while a CCD (AVT Stingray) images the particles in the bright field for user operation.

To maximise the available switch rate, the desired array of trap positions is communicated to the FPGA prior to running the experiment. As such, the total number of positions is limited by the onboard memory of the FPGA; in our configuration 1000 trap positions can be stored and cycled. By loading in successive trap positions that are only slightly displaced from one another, it is also possible to generate one or more oscillating traps, although the available frequencies of oscillation are limited by storage capacity and trap switching time. While in this paper we utilise this method to perform sinusoidal oscillation, any arbitrary movement of traps (such as triangle waves, step functions etc.) is possible.

Due to a combination of trigger timing, bandwidth limitations of the PSD and delays in DFS signal construction and propagation of the acoustic wave across the 2D-AOD, a trap switch is followed by a signal dead-time of approximately 60 \textmu s (Figure \ref{fig:apparatus} b) during which no position sensing can be performed. The maximum viable switch rate is determined by this dead time, as usable data can only be gathered 60 \textmu s after a switch trigger; we therefore use a switch time of 75 \textmu s. 

To gather viable data, the PSD is over-sampled by the FPGA at 200 kHz which results in a large volume of position samples within this dead time and some remaining usable data. We then down-sample the gathered data to exclude dead-time samples by using a virtual lock-in amplifier onboard the FPGA, which synchronises the binning and switch triggers. However, this method produces asynchronous clumps of single particle data separated in time by both large dead-times and periods where other particles were being probed. To allow computation of useful statistics (autocorrelations and cross-correlations) we further downsample the data to a rate of one data point per trap switch. As such, the effective data rate ($f_e$) of the sensing apparatus depends on the switch rate and number of traps; 
\begin{equation*}
f_{e} = \dfrac{1}{N_{traps}\Delta t}
\end{equation*}
For two traps at switch time of $ \Delta t = $ 75\textmu s this data rate is is 6.7 kHz.

Absolute calibration of each trapped particle is performed \textit{in situ} by utilising the fast switching of the 2D-AOD. In brief; particles are displaced by the optical trap a known distance using the 2D-AOD. The trap quickly switches back to the central position and the PSD response recorded; this is repeated for different distances to compute the PSD response as a function of particle position. Each trapped particle is calibrated in sequence, with the trapping laser periodically re-visiting the other particles to maintain localisation during calibration. Using this \textit{in situ} calibration, our detection method is robust to the size and shape of trapped particles and to an extent the interference of scattering laser light from adjacent particles. This provides an attractive potential for investigations of assemblies of non-uniform or non-spherical colloids \cite{kavre2015hydrodynamic}. In addition, as our calibration is performed separately for each trap position, it is robust to any minute angular dependencies in trapping efficiency or micron-scale differences in the trapping environment such as domain boundaries or distortions in the back focal plane interference pattern for traps a large distance from the optical axis of the objective.

Position determination for sinusoidally oscillating particles requires an extra step; the calibrated measurements of position relative to trap-centre are offset by the known position of the sinusoidal trap at time of sampling to yield measurements in the laboratory frame.
\section{Correlated Motion of Adjacently Trapped Particles}
\label{sec:double}
We utilised our time-multiplexed system to measure the correlations in motion of arbitrary patterns of adjacently trapped microspheres. We first considered two adjacent particles, for which the hydrodynamic interactions have an analytic solution that has been experimentally validated \cite{Meiners}, as a proof-of-concept for our system. We then examined several computational predictions for systems of three adjacent particles \cite{herrera2013hydrodynamic} in the short-time domain.
\subsection{Two Microspheres} 
Meiners and Quake \cite{Meiners} derived and experimentally showed that two adjacent micro-spheres in identical continuous-power traps will undergo motion that is anti-correlated at short time-scales as a consequence of their hydrodynamic interaction and relaxation within the optical trap \cite{polin2006anomalous}.

\begin{figure}
\centering
\includegraphics[scale=0.6]{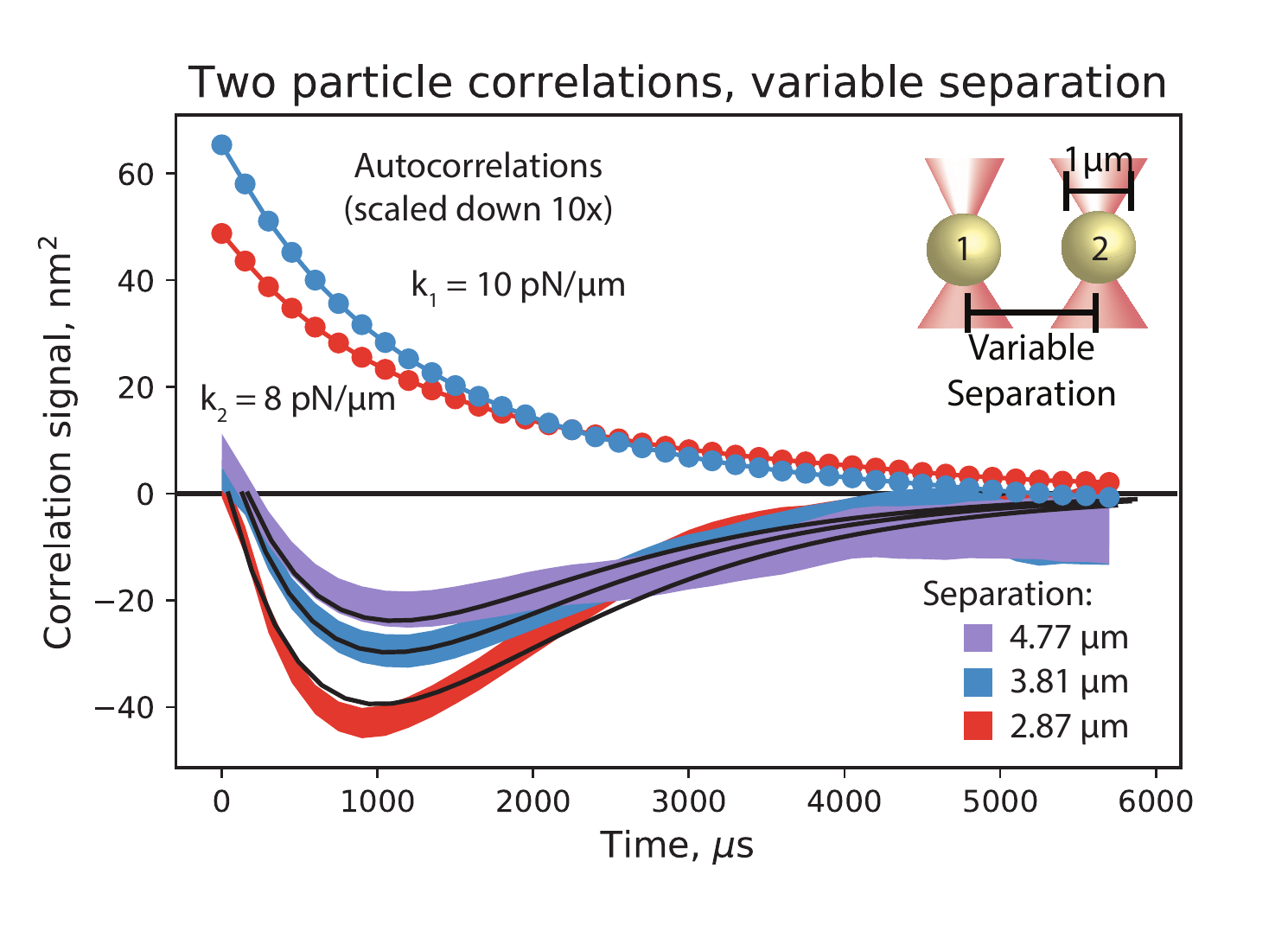}
\caption{Auto and cross-correlation measurements of two adjacently trapped 1\textmu m diameter spheres in a time-multiplexed system. Cross-correlation shows strong agreement with theoretical predictions from Eqn \ref{eqn:Berut}, with the only free parameter an x-axis offset. From this, we conclude that measurements using time-sharing retain validity when compared to continuous power systems. Auto-correlations for each trap are different due to differing trap stiffnesses ($k_1 = 1\:\mathrm{pN/\mu m}$, $k_2 = 8\:\mathrm{pN/\mu m}$), owing to diffraction efficiency changes in the AOD .} 
\label{fig:twotrap}
\end{figure}

As our system possesses small inhomogeneities between traps (due to the efficiency of the 2D-AOD depending slightly on the desired angle of deflection) our theoretical predictions utilise a modified formulation of Meiners and Quake's formulation, following Beirut's \cite{Berut2015} expression of the cross correlation between adjacent particles in traps of different stiffnesses $k_1$ and $k_2$: 
\begin{equation}
\label{eqn:Berut}
\braket{x_1(t)x_2(0)} = \dfrac{\epsilon k_b T}{\kappa}\big(e^{-(k_1+k_2-\kappa)t/2\gamma}-e^{(k_1+k_2+\kappa)t/2\gamma}\big)
\end{equation}
where:
\begin{equation*}
\kappa = \sqrt{k_1^2 - 2k_1k_2 + k_2^2 + 4\epsilon^2k_1k_2}
\end{equation*} 
Here, $T$ and $\gamma$ are the system temperature and Stokes' drag coefficient respectively.
The coupling coefficient, $\epsilon$ for two particles of radius $a$ and centre-centre separation $r$ is $\epsilon = \frac{3a}{2r}$ \cite{Berut2015}.
\\\\
The stiffness of each optical trap was measured using both the equipartition theorem \cite{neuman2004optical} and power spectra via the fluctuation-dissipation theorem \cite{berg2004power}. Theoretical predictions and experimental results are compared in Figure \ref{fig:twotrap} for particle separations of 2.87, 3.81 and 4.77 \textmu m. We find good agreement between theoretical predictions and our experimental measurements.

These results show that it is possible to use time-multiplexed optical tweezers to achieve a sensing capacity comparable to dual-beam continuous power apparatus. In addition, the multiplexing hardware used for this system allow calibration method substantially simpler (and quicker) than the stage driving method conventional to single-laser dual beam optical tweezers \cite{Meiners}\cite{visscher1996construction}.

\subsection{Three Microspheres}
Following this proof of concept, we examined theoretical predictions for the motion interactions between three adjacent colloids by Herrera-Velarde et al. \cite{herrera2013hydrodynamic}. Although the effective bandwidth of our system is reduced by the addition of a third trap (6.7 kHz to 4.5 kHz), we are still able to investigate the predicted colloidal behaviour at low time-scales. 

\begin{figure}
\centering
\includegraphics[scale=0.6]{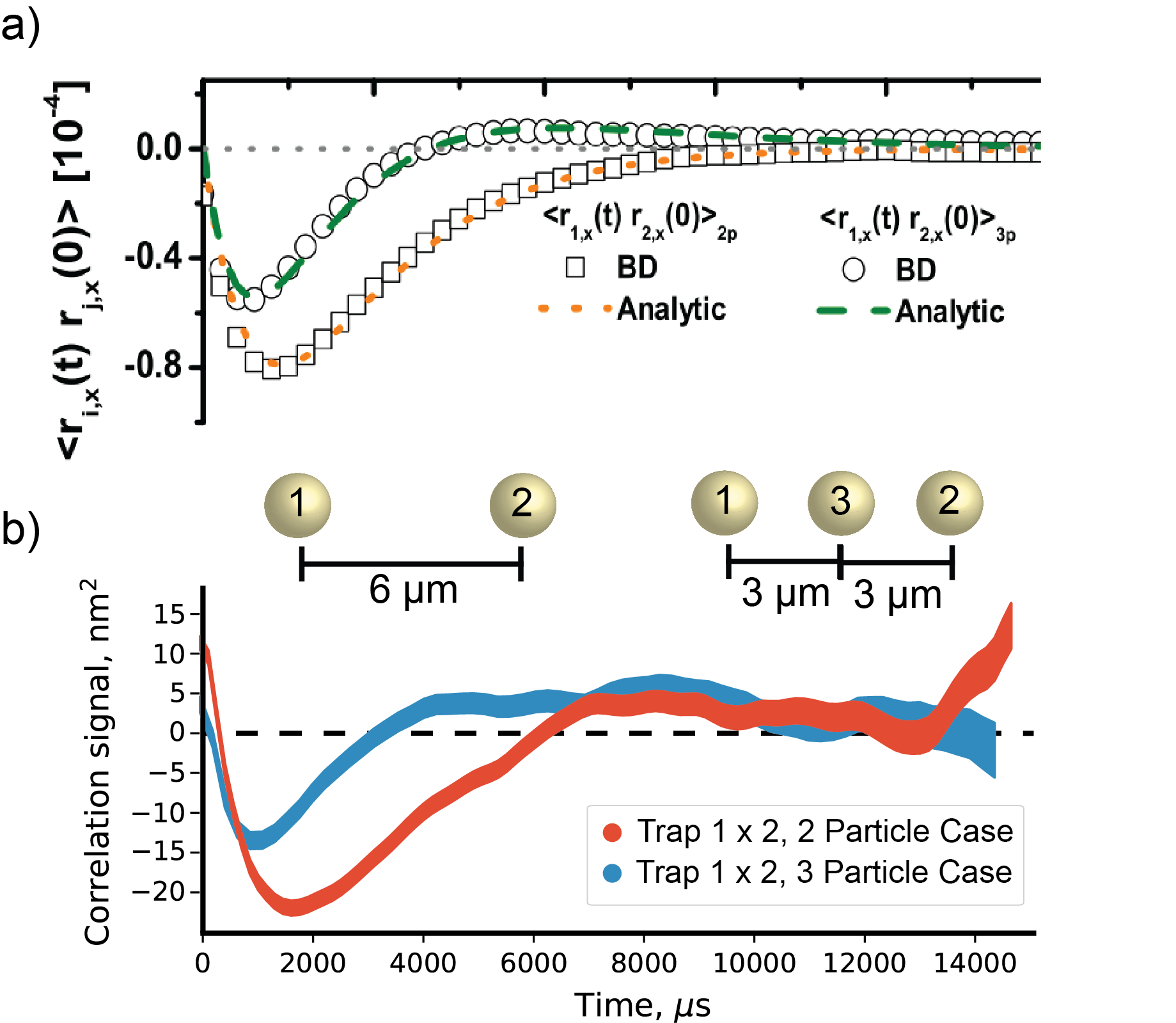}
\caption{Above: Simulated cross correlations between two particles with/without an intermediate particle conducted by Herrera-Velarde et al. \cite{herrera2013hydrodynamic} (figure reproduced). Below: experimental measurements of cross correlations between two 1 \textmu m spheres at a separation of 6 \textmu m, with the inclusion of an intermediate particle. Strong agreement is found between simulation and experiment; the intermediate particle reduces both the magnitude of anti-correlated motion and changes the position of the minima.}
\label{fig:threetrapmiddle}
\end{figure}

Herrera-Velarde et al. theorise that the magnitude of hydrodynamic coupling of two adjacent micro-spheres at will be lessened by the presence of a third micro-sphere in an intermediate position. In addition, they predict that the temporal position of the correlation minima will shift to a lower time (Figure \ref{fig:threetrapmiddle} a, replicated from \cite{herrera2013hydrodynamic}). To test this, we observed correlated motion between two 1 \textmu m spheres at a separation of 6 \textmu m, before inserting a third particle at the midpoint between the two and re-measuring the correlations of the outer colloids. Our results (Figure \ref{fig:threetrapmiddle} b) are as Hellera-Velarde et al. predict; a diminished and shifted anti-correlation upon the addition of an intermediate particle.

We further investigated the colloidal hydrodynamic behaviour as the central particle was moved; either horizontally (closer to one of the particles) or vertically (forming a triangle). In both cases, Herrera-Velarde et al. predict small changes in the magnitude of anti-correlation and the position of correlation minima. Unfortunately, these changes were within the error bounds of our measurements and were unable to be verified experimentally. 

Extensions to this system for trapping and tracking of larger groups of colloids (such as chains \cite{polin2006anomalous} or rings \cite{di2007eigenmodes}) is methodologically simple, but presents several inhibiting factors. Firstly, the effective data rate of the system decreases as more optical traps are added; for our switch rate of 75 \textmu s, $f_{e}$ decreases to below 1 kHz with 14 or more `simultaneous' optical traps. Secondly, increasing the number of trapped colloids increases the time interval between individual trap activations, which would lead to substantial deviations between the time-sharing and continuous power cases \cite{ren2010monte}. Fortunately, the maximum number of simultaneous particles can be extended simply by improving the possible switch time, which in our case is a combination of AOD delay and low sensor bandwidth.

However, an unavoidable limitation inherent in the time-sharing method is the asynchronous nature of data gathered. In a two particle system, this manifests simply as a small uncertainty in cross-correlation data and minimally affects results. For larger colloidal groups however, the time between samples for a particular pair of particles can become quite large. This reduces the usefulness of correlative measurements for any particles far apart in the multiplexing order.  We therefore suspect that for investigations involving large colloidal groups, time sharing is most appropriate for examining nearest-neighbour interactions, where these temporal deviations are the smallest. 
\section{Sinusoidal Driving of Adjacently Trapped Particles}
Using our time-multiplexed system, active driving of one or more colloids was achieved by successively displacing the position of one trap during switching. We explored the utility of this manipulation by conducting hydrodynamic driving of a single static particle with one or more sinusoidally oscillating colloids. Similarly to Section \ref{sec:double}, we first replicated an existing experiment to validate that our system imposed little errors on measurements of particle motion; Hough and Ou-Yang's driving of a single sphere with an identical trapped sphere \cite{HoughLA2002Cmot}. 

For a trap of stiffness $k_1$ sinusoidally oscillating at a frequency of $w$ and amplitude of $A$, the amplitude of particle oscillation $(X_1)$ can be determined from the real component of the hydrodynamic self-response tensor ($\chi_{11}$)  \cite{HoughLA2002Cmot};
\begin{align*}
X_1 &= Ak_1*\operatorname{Re}(\chi_{11})  \\
&= \dfrac{Ak_1}{(s_+ + s_-)k_1\tau_1}\Big[\dfrac{(k_2/k_1+\tau_1 s_+)s_+^2}{w^2+s_+^2} - \dfrac{(k_2/k_1+\tau_1 s_-)s_-^2}{w^2+s_-^2}\Big]
\end{align*}
An adjacent trapped particle (of stiffness $k_2$) will be perturbed by the displacement of fluid from the oscillating particle, exhibiting sinusoidal motion of a reduced amplitude. The amplitude of this motion ($X_{2}$) can be calculated from the real component of a cross-response tensor $\chi_{12}$ \cite{HoughLA2002Cmot};
\begin{equation*}
X_{2} = Ak_2*\operatorname{Re}(\chi_{12}) = \dfrac{Ak_2\epsilon}{(s_+ - s_-)k_1\tau_2}\Big[\dfrac{s_-^2}{w^2+s_-^2} - \dfrac{s_+^2}{w^2+s_+^2}\Big]
\end{equation*}
Here, $\tau$ is the autocorrelation time for the trap in question ($\tau_n = \frac{1}{\gamma k_n}$) and $s_+$ and $s_-$ are the poles of the response tensor, given by \cite{HoughLA2002Cmot};
\begin{equation*}
s_{\pm} = \dfrac{-1+k_2/k_1 \pm \sqrt{(1-k_2/k_1)^2 + 4 k_2/k_1 \epsilon^2}}{2\tau}
\end{equation*} The magnitude of sinusoidal response ($X_1, X_2$) was determined using the standard deviation: $X_n = \sqrt{2}\sigma(\mathbf{x_n})$, which functions as a useful measure of the mean amplitude of oscillation.  The endemic Brownian motion of the particles was accounted for by first keeping the traps in a static condition, determining a baseline Brownian deviation and subtracting this baseline from subsequent measurements of $X_1$ and $X_2$. 

\begin{figure}[h]
\centering
\includegraphics[scale=0.6]{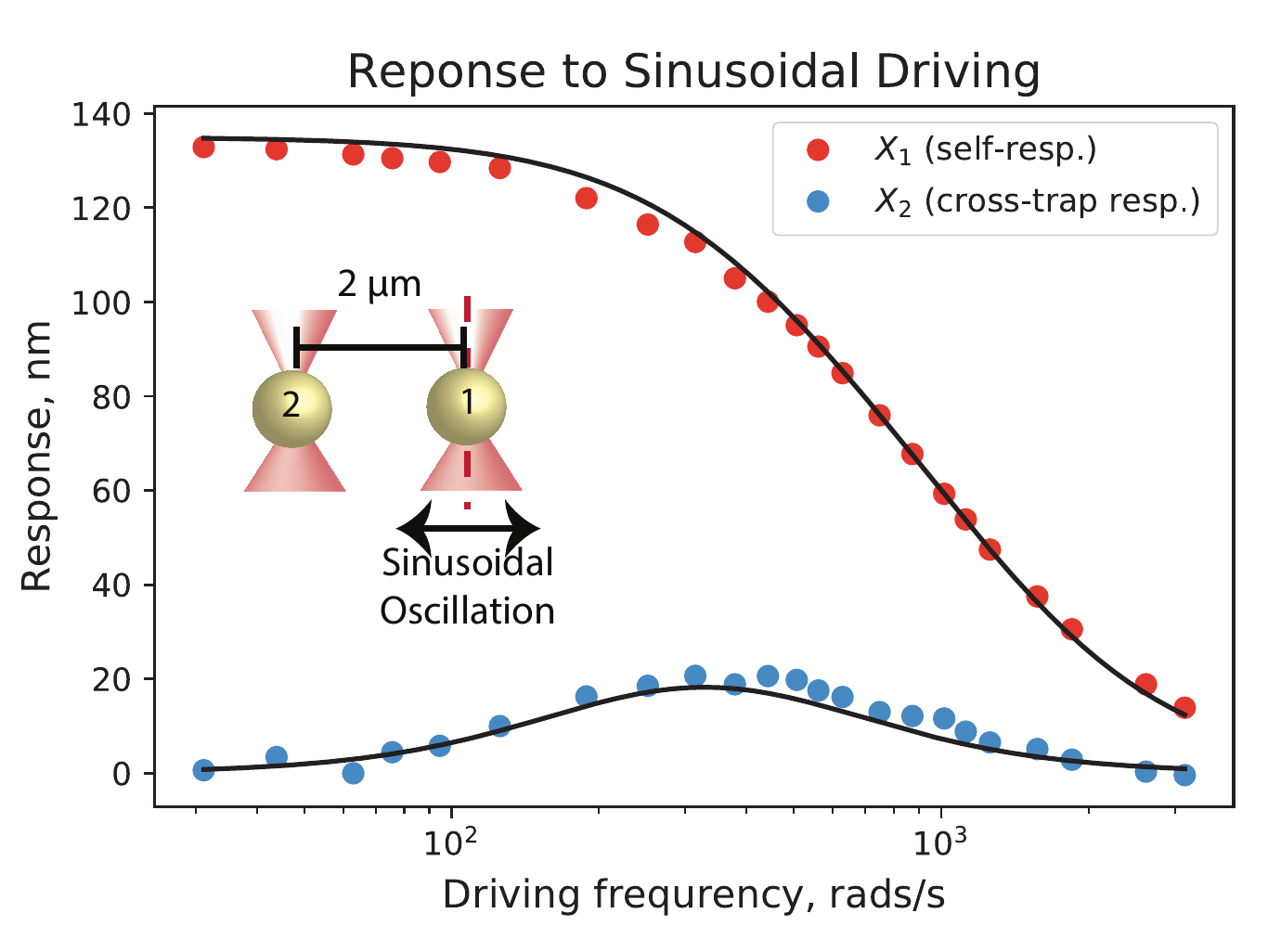}
\caption{Response to sinusoidal driving of trap 1 for a particle within the trap (driving) and in an adjacent trap (driven). We find (as demonstrated in \cite{HoughLA2002Cmot}) a resonance peak for the driven particle, and a decrease in response for the driving particle with increasing frequency due to hydrodynamic drag. The response of the particle was experimentally determined by comparing the rms deviation with the non-oscillating case. Solid lines show theoretical predictions for response.} 
\label{fig:sinusoid1}
\end{figure}

Figure \ref{fig:sinusoid1} shows measured response and theoretical predictions for two particles separated by a distance of 2 \textmu m with oscillation amplitude 0.2 \textmu m. We find strong agreement between our results and the theory, further validating time-sharing as a suitable technique for multi-colloidal optical tweezer investigations.

Following this success, we measured the response of a central particle with two adjacent oscillating particles, in a manner similar to \cite{leoni2009basic}. The two oscillating traps were placed a mean distance of 2 m from a central trap, and oscillated both in phase and out of phase. We find similar self-response curves for the oscillating particles, with minor discrepancies due to diffraction efficiency differences within the AOD.
\begin{figure}
\centering
\includegraphics[scale=0.5,left]{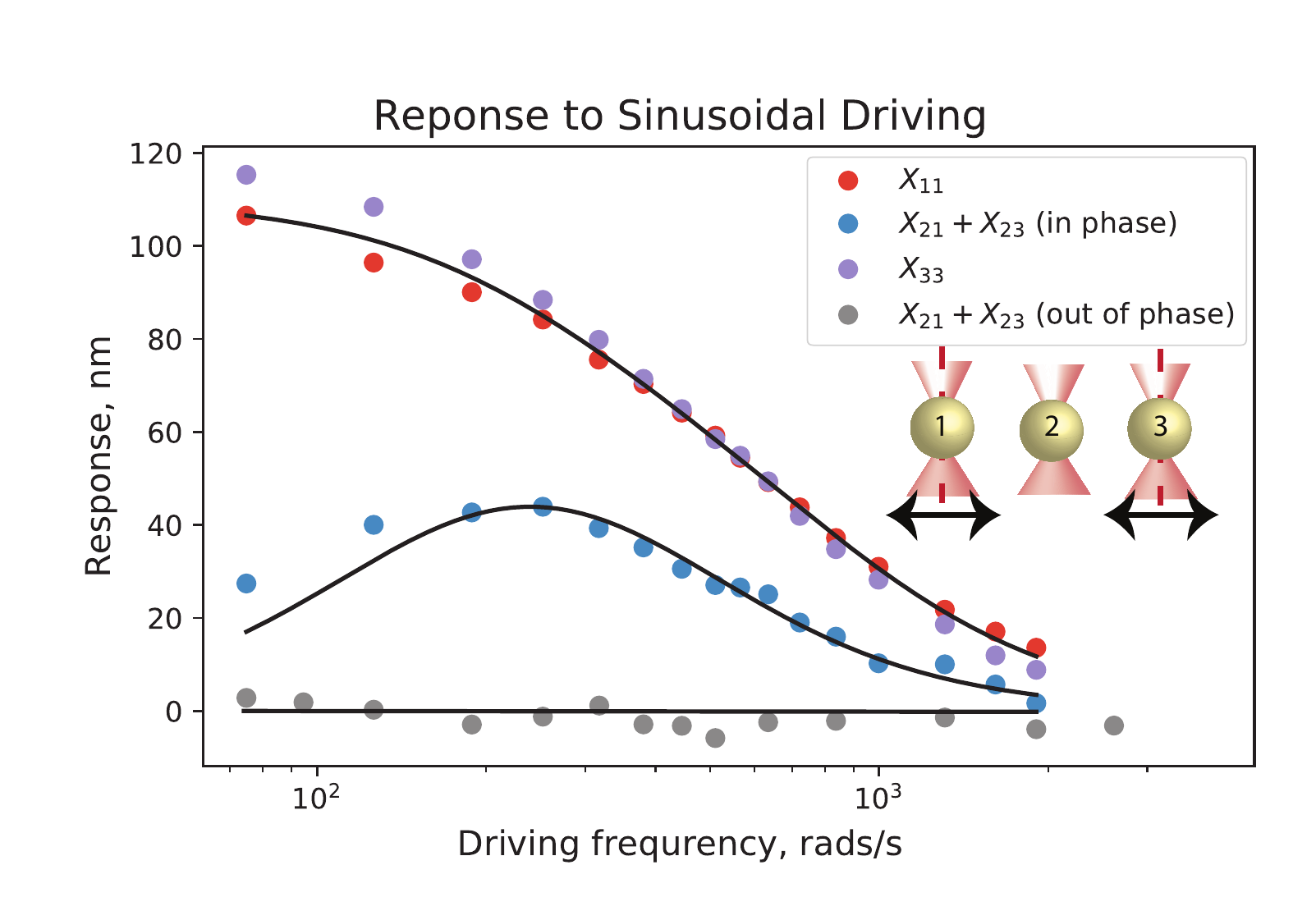}
\caption{Response of a central particle (trap 2) to sinusoidal driving from either side, with solid lines indicating relevant theoretical predictions. Self response ($X_1 \& X_3$) are as expected. The resonance of the doubly-driven particle is approximately double the single-driver case when the two resonators are in phase, and effectively zero when the two are out of phase. Utilising three traps over two has shifted the position of the resonance by reducing the stiffness of each trap.} 
\label{fig:sinusoid2}
\end{figure}
Figure \ref{fig:sinusoid2} shows the driven particle response for in phase and out of phase oscillations. When the sinusoids are in phase, the response of the driven particle is much larger than in the single oscillator case. The response for two oscillators out-of-phase is almost entirely absent, suggesting that the influence of each oscillator has destructively interfered. We make theoretical predictions for the composite response for the in phase case by simply adding the real components of the response tensors:
\begin{equation*}
X_2 \approx A_1 k_1 \operatorname{Re}(\chi_{21}) + A_3 k_3 \operatorname{Re}(\chi_{23})
\end{equation*}
Where $\chi_{23}$ is the response of the central trap (2) to the second driving trap (3), and is otherwise identical to $\chi_{21}$. We find a moderate agreement between this basic prediction and observations, noting that no interaction effects between oscillators were taken into account.

Beyond simplistic linear arrangement of oscillators, this system can easily be extended to examine active colloidal configurations of interest in biological sciences. Non-linear arrangements such as ring-oriented spherical colloids \cite{damet2012hydrodynamically}\cite{cicuta2012patterns}, or oscillations of non-spherical particles \cite{kavre2015hydrodynamic} are achievable at observational time-scales faster than conventional camera imaging.  
\section{Conclusion}
Time-multiplexed optical tweezers have the potential to allow powerful investigations of multi-colloidal hydrodynamics. In this paper, we have demonstrated and validated a simple stroboscopic setup capable of trapping of arbitrary configurations of colloids, sinusoidal driving and position sensing in the kHz regime. Our multiplexing-enabled calibration mitigates several of the issues associated with time-sharing of optical traps using Accousto-Optic Deflectors, and provides robustness against colloidal shape distortions and local inhomogeneities.

We expect this method to be useful for investigations into short-timescale multi-colloidal hydrodynamics, where interaction effects are beyond the sample rate of conventional camera based position sensing. Nominal extensions of this system to cover more complex arrangements of static or active colloids are easily achievable with small software changes. Increases in potential driving frequencies or improvements in the effective data rate will require a reduction of system dead-time, which can be achieved through improvements to the speed of the AOD-DFS and photodiode acquisition, pushing both position sensing and colloidal driving into the 10s of kHz. 

\bibliography{references}
\end{document}